\pdfoutput=1

\documentclass[11pt]{article}
\usepackage{natbib,bm}
\bibpunct[; ]{(}{)}{,}{a}{}{;}

\usepackage[ruled]{algorithm2e}
\usepackage{authblk}
\usepackage[utf8]{inputenc}
\usepackage{gensymb}
\usepackage{hyperref}
\usepackage{graphicx} 
\usepackage{subcaption} 
\usepackage{amsmath, amssymb, bm}
\usepackage{units} 
\usepackage{multirow}
\usepackage[a4paper]{geometry}
\usepackage[labelfont=bf]{caption}
\usepackage{float} 
\usepackage{setspace}
\providecommand{\keywords}[1]
{\small	\textbf{Keywords: } #1 }

\usepackage[flushmargin]{footmisc}

\def\boldfacefake#1{\kern-4pt
    \hbox{ \mathsurround=0pt
    \hbox to 0.4pt{$#1$\hss}\hbox to 0.4pt{$#1$\hss}\hbox {$#1$}}}

\newcommand{\be}{\begin{eqnarray}}
\newcommand{\ee}{\end{eqnarray}}
\newcommand{\ba}{\begin{eqnarray*}}
\newcommand{\ea}{\end{eqnarray*}}
\newcommand{\bc}{\begin{center}}
\newcommand{\ec}{\end{center}}
\newcommand{\btab}[1]{\begin{tabular}{#1}}
\newcommand{\etab}{\end{tabular}}

\newcommand{\reals}{\mbox{\rm I\kern-.20em R}}
\newcommand{\sreals}{\mbox{\small \rm I\kern-.20em R}}

\title{A robust multivariate, non-parametric outlier identification method for scrubbing in fMRI}

\author[1]{Fatma Parlak \thanks{Corresponding author: Fatma Parlak, fparlakstat@gmail.com}} 
\author[1]{Zahid Zeshawn}
\author[1]{Damon D. Pham}
\author[1]{Amanda F. Mejia}
\date{}

\affil[1]{Department of Statistics, Indiana University, Bloomington, IN, USA}

\begin{document}

\maketitle

\begin{abstract}
Functional magnetic resonance imaging (fMRI) data contain high levels of noise and artifacts due to head motion, scanner instabilities, and other sources. To avoid contamination of downstream analysis, fMRI-based studies must identify and remove these sources of noise prior to statistical analysis. One common approach is the ``scrubbing'' or ``censoring'' of fMRI volumes that are thought to contain high levels of noise. However, existing scrubbing techniques are based on subject head motion measures or ad hoc measures of signal change. Here, we consider scrubbing through the lens of outlier detection, where volumes containing artifacts are thought of as multidimensional outliers. Robust multivariate outlier detection methods have been proposed using robust distances, which are related to the Mahalanobis distance. These robust distances have a known distribution when the data are i.i.d. Gaussian, and that distribution can be used to determine an appropriate threshold for outliers. However, in the fMRI context, we observe clear violations of the assumptions of Gaussianity and independence. Here, we develop a robust multivariate outlier detection method that is applicable to non-Gaussian data. The objective is to obtain threshold values to flag outlying volumes based on their robust distances. We propose two threshold candidates that embark on the same two steps, but the choice of which depends on a researcher's purpose (i.e., greater data retention vs. greater sensitivity). The two main steps of our procedure are: (1) dimension reduction and selection to identify primarily artifactual latent directions in the data; (2) robust univariate outlier imputation to remove the influence of outliers from the distribution; and (3) estimating a threshold for outliers based on the upper quantile of the outlier-free distribution of RD. We propose two threshold choices. The first threshold is an upper quantile (e.g., $99^{th}$) of the empirical distribution of robust distances obtained from the imputed data. The second threshold is an estimation of the upper quantile of the robust distance distribution employed by a nonparametric bootstrap to account for uncertainty in the empirical quantile. We compare our proposed approach with existing approaches for scrubbing in fMRI, including motion scrubbing, data-driven scrubbing, and existing multivariate outlier detection methods based on restrictive parametric assumptions.

\end{abstract}

\section{Introduction}
Functional magnetic resonance imaging (fMRI) is a non-invasive technique that can be used to localize task-specific active brain regions and assists in predicting psychological or disease states \citep{lindquist2008statistical}. During neuronal activity, the brain hemodynamics alter due to an increase in the blood oxygenation level in the brain cells. These changes in the brain are captured in a magnetic resonance imaging (MRI) scanner. These MR images are collected over the time ($2 \sim 60$ min.) of the experiment and consist of $\sim100,000$ evenly sized cubes known as voxels, which collectively represent an individual’s whole brain. Therefore, fMRI data are a form of high-dimensional data containing the received signals from each voxel at each time point. An acquired image from all voxels at one time point is called a $\textit{volume}$. 

To employ artifact-contaminated fMRI data reduces the quality of the results and influences the statistical result by reducing the signal-to-noise ratio (SNR) and violating common statistical assumptions. A low SNR is one of the shortcomings of fMRI that makes it more difficult to identify active brain regions associated with an activation task because these artifactual signals might mask the real brain signals. Artifacts may be either participant-related (such as, head movements, eye movements, breathing, and heartbeats \citep{friston1996movement, beauchamp2003fmri, frank2001estimation, kruger2001physiological}) or equipment-related (spikes, scanner drift). The identification of artifact-contaminated volumes before data analysis, is crucial. 

In this work, we propose a robust outlier detection approach for identifying artifact-contaminated fMRI volumes. Our approach considers the high-dimensional, auto-correlated, and non-Gaussian aspects of fMRI data. Standard outlier detection methods fail in high-dimensional data. However, in many cases dimension reduction techniques can be used, after which existing multivariate outlier detection methods can be applied. The literature provides various distance-based outlier detection methods for multivariate data. One of the oldest is Mahalanobis distance \citep{mahalanobis1936generalised} which is based on the sample mean and sample covariance. However, both the center and scaling factors are obtained by using all observations and therefore may be influenced by outliers. The influence of outliers on the sample mean and covariance can cause ``masking" and ``swamping" effects \citep{rousseeuw1990unmasking}. The masking effect is failure to identify true outliers, while the swamping effect is to flag non-outliers as outliers. As such, Mahalanobis is a non-robust measure.  

\cite{rousseeuw1990unmasking} proposed the use of robust minimum covariance determinant (MCD) estimates of mean and covariance in place of the conventional ones to produce a robust distance (RD) measure. MCD estimates are calculated by splitting the data into two subsets, one of which contains the observations located close to the center of the data which is unlikely to represent outliers, is used to estimate the location and scale parameters robustly.  

\cite{hardin2005} derived the theoretical distribution of MCD-based RD for Gaussian distributed data. An upper quantile of that distribution can be used to identify outliers. For example, using the $99^{th}$ quantile of the theoretical distribution, on average $1\%$ of non-outlying observations will be flagged as outliers along with the others. Unfortunately, the empirical distribution of RD has often been observed to deviate from the theoretical distribution. Previous work employing MCD-based RDs attempted ad-hoc approaches to improve the distributional fit, for example median matching \citep{mejia2017pca, filzmoser2008, maronna2002}. Some reasons for this deviation might be due to violation of assumptions of independence, identical, and Gaussianity. fMRI data typically violate these key assumptions, in particular those of Gaussianity and independence. Therefore the use of Hardin \& Rocke approach fMRI data can cause high false positives and incorrectly scrubbing the beneficial volumes as artifactual.

Here, we propose a novel non-parametric robust multivariate outlier detection method that is applicable to fMRI data. We consider two approaches to threshold outliers: an empirical quantile calculation of the robustly transformed data's RDs and an upper quantile estimation of original data's RDs via a non-parametric bootstrap. These methods can be applied to any type of low-dimensional data, so we assume that the fMRI data has undergone dimension reduction using previously proposed techniques \cite{pham2023less}. First, we define the two MCD subsets. Second, we apply a robust univariate outlier imputation proposed by \cite{raymaekers2021transforming} to mitigate the influence of outliers on the procedure. Finally, we identify outliers based on the threshold values. One way of thresholding the RDs is to calculate an empirical quantile of the RDs via robustly transformed data. Another way is to employ a non-parametric bootstrap within each subset to estimate the distribution of RDs and the quantile for thresholding outliers. These thresholds are applied to the RDs of the original data to identify outliers. 

The remainder of this paper is organized as follows. In Section \ref{sec:methods}, we introduce our proposed method and compare it with \cite{hardin2005}'s theoretical approach using simulated data and two toy fMRI datasets. In Section \ref{sec:dimred}, we apply a recently proposed dimension reduction and selection approach to the fMRI data. In Section \ref{sec:RD}, we briefly describe how MCD subset selection and RD calculation are employed from a recently proposed method. In Section \ref{sec:distRD}, we explain and illustrate the consequences of utilizing Hardin \& Rocke's approach for the distribution of RDs in terms of identifying outliers in both fMRI data and simulated fMRI data. In Section \ref{sec:bootRD}, we describe our proposed approach to threshold the distribution of RDs to identify outliers. In Section \ref{sec:uoi}, we describe our robust univariate outlier imputation algorithm. In Section \ref{sec:estQ}, we describe the estimation of an upper quantile of the RDs, considering two quantile calculations to threshold RDs. In Section \ref{sec:EDA}, we apply our method to fMRI data
from the Human Connectome Project (HCP) 42-subject retest \footnote{http://humanconnectome.org} and compare it with existing methods \citep{afyouni2018insight, smyser2010longitudinal, pham2023less, power2012, power2014}. In Section \ref{sec:discussion}, we briefly summarize the results and discuss limitations, suggesting future work. 

\section{Method}
\label{sec:methods}
Here, we propose a novel, robust, high-dimensional outlier detection method that can be used to identify artifact-contaminated fMRI volumes. Since neuronal activity accounts for only a small portion of variance in fMRI data, temporal spikes or outliers in the BOLD signal are assumed to be of artifactual origin. Our approach consists of four steps, which are described in the following subsections. The first step is to reduce the dimensions and select high-kurtosis components, which are likely to represent artifactual volumes. For low-dimensional data, this step can be skipped. The second step is to compute a robust distance (RD) measure. The third step, which is our novel contribution, is to estimate the null distribution of RDs non-parametrically and obtain an upper quantile of this distribution to serve as an outlier detection threshold. The fourth step is to apply this threshold to identify artifactual volumes.

\subsection{Dimension reduction and selection}
\label{sec:dimred}
Here, we adopt a dimension-reduction and selection approach that we recently proposed and validated \citep{pham2023less}. Briefly, we reduce the dimensionality of the data by applying independent component analysis (ICA) and then select high-kurtosis components likely to represent artifactual patterns in the data. This approach improves upon the PCA leverage method proposed by \cite{mejia2017pca} and selects certain components based on kurtosis. As the use of ICA is a popular tool for dimension reduction and for identifying spatially independent components in fMRI, which can be of neuronal or artifactual origin, these artifactual noises can be identified and removed in fMRI datasets by ICA using ICA-FIX \citep{griffanti2014} and ICA-AROMA \citep{pruim2015}. \cite{pham2023less} comprehensively compared the effectiveness of this novel ICA-based scrubbing approach, termed "projection scrubbing," with data-driven (DVARS \cite{afyouni2018insight, smyser2010longitudinal}) and head-motion-based scrubbing techniques \citep{power2012, power2014}. They found that projection scrubbing tended to produce more reliable and valid estimates of functional connectivity (FC), while retaining much more data than motion scrubbing. In this work, we, therefore, adopt the proposed dimension-reduction and selection approach, described next, while introducing an improved technique for identifying artifactual volumes.

Consider an fMRI dataset, ${\bf{Y}}_{T \times V}$,  where $T$ is the duration of an fRMI experiment, and $V$ is the total number of voxels of the brain ($T \ll V$). ICA decomposes ${\bf{Y}}$ into a spatial source signal matrix $({\bf{S}})$ containing independent components (ICs) and a temporal mixing matrix $({\bf{A}}_{T \times Q})$ containing the temporal activation profile of each IC. That is, ${\bf{Y}}={\bf{A}} \ {\bf{S}} + {\bf{E}}$. ${\bf{A}}$ can be considered a dimension-reduced version of ${\bf{Y}}$ along the spatial dimension. While the ICs of ${\bf{S}}$ may represent signal or noise, artifacts tend to appear in burst noise causing high spikes in the corresponding time courses. Therefore, the columns of ${\bf{A}}$ related to artifactual ICs are more likely contain extreme values. Kurtosis is an indicator of the presence of potential outliers. Since outliers can affect the tail of the distribution to be heavy-tailed, a higher kurtosis value is an indication of the existing outliers. Therefore, the components having \textit{high kurtosis} value are selected from the dimension-reduced data $({\bf{A}}_{T \times Q})$. Suppose we select the $K$ high kurtosis components within the $Q$ independent components. The resultant matrix, ${\bf{X}}_{T \times K}$, consists of the columns of $({\bf{A}}_{T \times Q})$ corresponding to the rows of ${\bf{S}}$ that are likely to represent artifacts. The kurtosis values are calculated as:

\begin{align}
    \text{Kurt} = \frac{1}{N}  \mathlarger{\sum}_{i=1}^{Q} \Bigg( \frac{a_{i} -\bar{a}}{s} \Bigg)^4 -3 , 
    \label{eqn:KURT}
\end{align}

for each component (${\bf{a}} = (a_1, a_2, ..., a_Q)$) of $({\bf{A}}_{T \times Q})$, where $\bar{a}$ and $s$ are the sample mean and standard deviation of $\bf{a}$, respectively. The goal of our method, described below, is to detect $K$-dimensional outliers among the $T$ observations of ${\bf{A}}$. These extreme observations can then be excluded from analysis via ``scrubbing" to avoid undue influence. 

\subsection{Robust distance calculation}
\label{sec:RD}

Let ${\bf{X}} = ({\bf{x}}_1, {\bf{x}}_2, ..., {\bf{x}}_n)^T$ represent data with $n$ observations across $p$ dimensions. The Mahalanobis distance (MD) of the $i^{th}$ observation ${\bf{x}}_i = ({x}_{i1}, {x}_{i2}, ..., {x}_{ip})^T$ is defined as
$\text{MD}({\bf{x}}_i) = \sqrt{({\bf{x}}_i - \hat{{\boldsymbol{\mu}}})^T \ {\hat{{\boldsymbol{\Sigma}}}}^{-1} \ ({\bf{x}}_i - \hat{{\boldsymbol{\mu}}})}$, where ${\hat{\boldsymbol{\mu}}} = ({\hat{\mu}}_1, \hat{\mu}_2, ..., \hat{\mu}_p)^T$  and $\hat{{\boldsymbol{\Sigma}}}_{p \times p}$ are the sample mean and sample covariance, respectively, across the $n$ observations. \cite{rousseeuw1990unmasking} proposed using minimum covariance determinant (MCD) estimators of the mean, $\hat{{\boldsymbol{\mu}}}_{MCD}$, and covariance, $\hat{{\boldsymbol{\Sigma}}}_{MCD}$, to mitigate the influence of outliers on the MD. These MCD estimates are obtained by choosing a subset of $h < n$ observations closest to the center of the distribution, and therefore unlikely to represent outliers. The MCD achieves its maximum breakdown point by choosing $h = \lfloor{(n+p+1)/2}\rfloor$ \citep{lopuhaa1991}. For $n$ observations ${\bf{x}}_{1}, {\bf{x}}_{2}, ..., {\bf{x}}_{n}$ with $p$ dimensions, the subset of $h$ observations, ${\bf{x}}_{i_1}, {\bf{x}}_{i_2}, ..., {\bf{x}}_{i_h}$, is chosen to provide the minimum possible determinant of the covariance among any subset. That is, letting $S_1 = \{i_1, i_2, ..., i_h\}$, $det ( \hat{{\boldsymbol{\Sigma}}}_{S_1}) \le det ( \hat{{\boldsymbol{\Sigma}}}_{S})$ for any set $S = \{k_1, k_2, ..., k_h\}$. We call the observations ${\bf{x}}_{i_1}, {\bf{x}}_{i_2}, ..., {\bf{x}}_{i_h}$ \textit{included} observations and the remaining observations \textit{excluded} observations. Let $S_2 = \{ 1,2, ...,n \} \backslash S_1$ index the excluded observations. The determination of the set $S_1$ requires an assessment of $\binom{n}{h}$ possibilities, which is computationally demanding as $h \approx \frac{n}{2}$. To overcome this computational challenge, we use an algorithm, \texttt{FastMCD}, developed by \cite{rousseeuw1999}. The MCD estimators, $\hat{{\boldsymbol{\Sigma}}}_{MCD}$ and $\hat{{\boldsymbol{\mu}}}_{MCD}$, are obtained based only on the included observations. Since MCD estimators have a breakdown point of $1-\frac{h}{n}$, the estimator of the mean and covariance remains robust as long as the proportion of outliers in the data does not exceed $1-\frac{h}{n}$. Using MCD estimates of the mean and covariance in the MD formula results in a robust distance (RD) metric. 
$$
\begin{aligned}
RD({\bf{x}}_i) &= \sqrt{({\bf{x}}_i - \hat{\boldsymbol{\mu}}_{MCD})^T \ {\hat{{\boldsymbol{\Sigma}}}_{MCD}}^{-1} \ ({\bf{x}}_i - \hat{{\boldsymbol{\mu}}}_{MCD})}
\end{aligned}
$$

\begin{figure}[H]
    \centering
    \includegraphics[height=0.6\textwidth, trim=0 0in 0in 0.195in, clip]{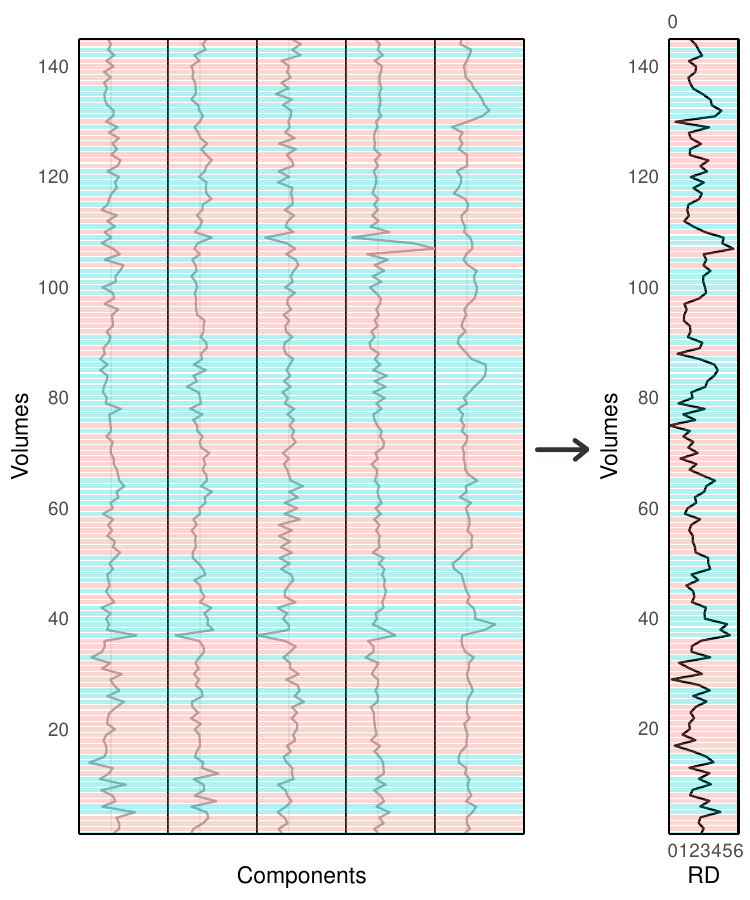} \\
    \includegraphics[width = 0.4\linewidth, trim=0 0in 0in 3.6in, clip]{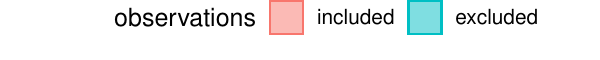}
    \caption{\small \textbf{Illustration of MCD-based robust distance calculation.} For illustration purposes, we chose $6$ components (out of 26) from the dimension-reduced and high-kurtosis components selected from fMRI data. Each component is a time series. The orange lines indicate included observations (volumes), while the turquoise ones indicate excluded observations. We obtain RDs of each observation in the time series based on the mean and covariance calculated from the included observations. As we can see from the left image, volumes with higher signal intensities are labeled as included, while volumes with lower signal intensities are labeled as excluded based on the MCD calculation. The right image shows the RDs of each volume across all six components. The excluded observations tend to have higher signal occurrences and higher RDs than the included observations. This figure visually illustrates the idea of expecting outliers in the excluded subset as it contains volumes that contain relatively higher signal intensities than the volumes in the included subset. The RD metric provides a single value for each volume that indicates the location of the volume relatively to the center of the data.}
    \label{fig:matrixRD}
\end{figure}

Figure \ref{fig:matrixRD} provides an illustration of the RD computation using toy fMRI data based on the publicly available Autism Brain Imaging Data Exchange (ABIDE) data \citep{di2014ABIDE}. These toy data are available in the \texttt{fMRIscrub R} package. The data consist of a single slice of a session represented by a $T \times V$ matrix , where $T = 145$ and $V = 4679$. This session was determined to be relatively free of artifacts based on visual inspection.  

\subsection{Theoretical distribution of MCD-based robust distance}
\label{sec:distRD}

To identify outliers based on RDs, it is necessary to know the distribution of RDs among non-outlying observations. An upper quantile of that distribution can be used as a threshold to identify outliers. \cite{hardin2005} derived the distribution of MCD-based RDs of i.i.d. Gaussian data. They proved that the RDs of excluded observations approximately follow a scaled $\textit{F}$ distribution. However, this theoretical result becomes invalid if any of the assumptions of Gaussianity, independence, or identical distribution are violated. As we show below, these assumptions are usually violated for fMRI data, making this theoretical result inapplicable, especially given that the chosen dimensions are selected based on having higher kurtosis than Gaussian data.  

\begin{figure}[H]
\centering
\begin{tabular}{ccc}
\begin{subfigure}{0.32\linewidth}
    \centering
    \includegraphics[width=0.9\textwidth, trim=0 0.5in 0in 0in, clip]{{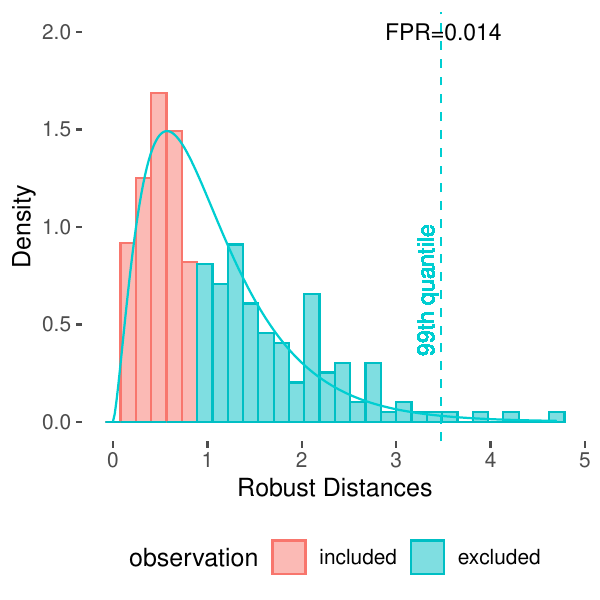}} \\
    \hbox{}
    \caption{i.i.d. Gaussian data}
\end{subfigure} &
\begin{subfigure}{0.32\linewidth}
    \centering
    \includegraphics[width=0.9\textwidth, trim=0 0.5in 0in 0in, clip]{{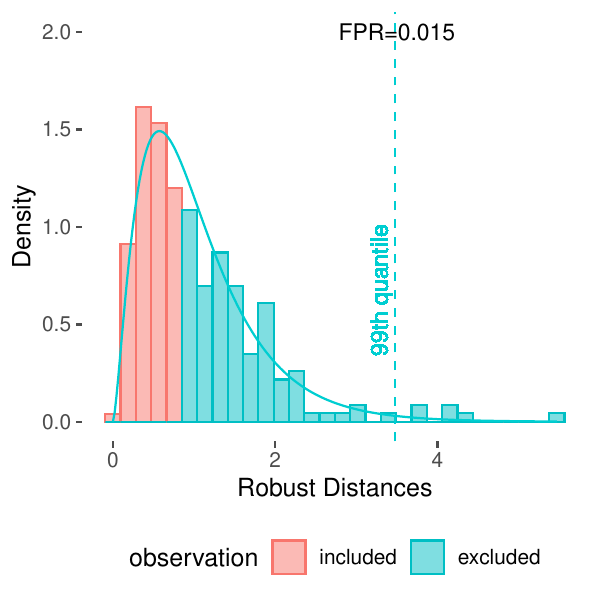}} \\
    \includegraphics[width = 0.9\linewidth, trim=0 0in 0in 3.6in, clip]{image/Figure3_Distributions_of_RD_from_ABIDE2.pdf} 
    \caption{AR(1) model, $\phi=0.4$}
\end{subfigure} &
\begin{subfigure}{0.32\linewidth}
    \centering
    \includegraphics[width=0.9\textwidth, trim=0 0.5in 0in 0in, clip]{{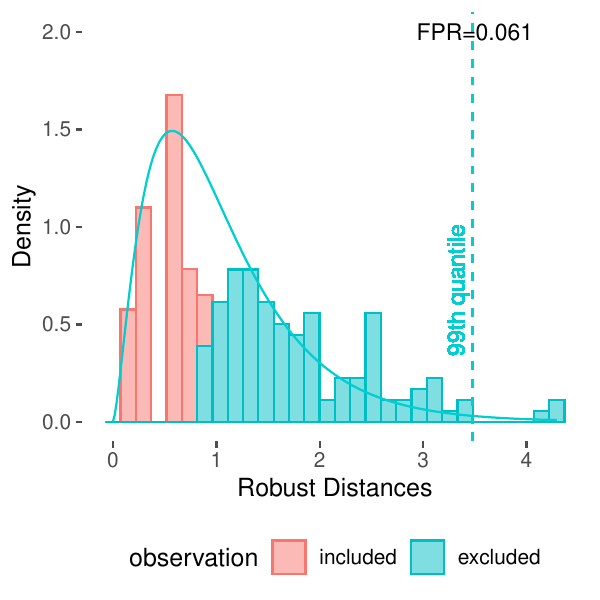}} \\
    \hbox{}
    \caption{AR(1) model, $\phi=0.9$} 
\end{subfigure}
\end{tabular} \\
    \caption{\small \textbf{Distributions of RDs in simulated Gaussian data with and without dependence and fit of the theoretical F distribution.} The vertical dashed lines indicate the 99th quantile of the F distribution proposed by \cite{hardin2005}. Reported false positive rates (FPR), written on top right of each panel, are obtained by averaging the replicated simulation with a nominal rate of $0.01$ for each setting $100$ times. (A) i.i.d Gaussian data are simulated. The mean FPRs is very close to the nominal rate of 0.01. (B) A first-order autoregressive (AR(1)) model with a correlation coefficient of $\phi=0.4$ is simulated. Even though the independence assumption of the theoretical method is violated, the average FPR does not show a significant change. (C) A first autoregressive (AR(1)) model with a correlation coefficient of $\phi=0.9$ is simulated. Minor violations of the independence assumption have negligible effects on the validity of the theoretical F distribution, but stronger dependence patterns may alter the distribution more substantially. }
    \label{fig:sim}
\end{figure}

In Figure \ref{fig:sim}, we visualize Hardin \& Rocke's theoretical result on three outlier-free and Gaussian simulated datasets. Because the excluded observations are a subset of the dataset with a distribution truncated at the $(1-\frac{h}{n})$-th quantile, both included and excluded observations are displayed on the histograms for ease of visualization. The scaled F distribution is displayed to show the empirical versus theoretical distribution fit of excluded observations' RDs. Observations are identified as outliers when they lie beyond the vertical lines at the $99^{th}$ quantile of the theoretical F distribution. We replicate these three datasets $100$ times with the same settings to compute the average false positive rate (FPR), which indicates the rates of observations being wrongly labeled as outliers. Since the $99^{th}$ quantile is used, an FPR near $1 \%$ is expected.

Figure \ref{fig:sim}a shows the RDs of the first dataset, which is generated from independent and identical Normal distribution. Recall that under these assumptions, the excluded observations' RDs follow a scaled F distribution. The second and third datasets, shown in Figure \ref{fig:sim}b and Figure \ref{fig:sim}c, respectively, are generated from a Gaussian first-order auto-regressive ($AR(1)$) model to reflect one feature of fMRI data: autocorrelation. To mimic typical fMRI data, we choose $\phi = 0.4$, a value that is commonly assumed for the temporal correlation of single-band fMRI data. For the third dataset, we choose $\phi = 0.9$, which imitates a more modern fMRI acquisition with fast temporal resolution. The theoretical F distribution fits reasonably well on the three histograms displayed in Figure \ref{fig:sim}. As expected, the $99^{th}$ quantile lies at the tail of the distribution of excluded observations’ RDs for each dataset, near the nominal level of $1 \%$. The average FPRs in the three scenarios are $1.3 \%$, $1.5 \%$, and $6.1\%$ respectively. The assumption of independence is violated in the second and third datasets and these FPRs suggest that the violation of the independence assumption has an effect on the validity of the theoretical F distribution.

While a distributional fit could be improved by accounting for the dependence through effective sample size, fMRI data also exhibit major deviations from the assumption of Gaussianity. The degree of dependence is known to vary dramatically across the brain \citep{Parlak2022-sz}. Therefore determining the effective sample size in fMRI data is non-trivial, and would need to be done in a robust manner to avoid the influence of outliers. Although the spatial autocorrelation pattern is accounted via bias correction, these correction strength also vary across the brain \cite{afyouni2019effective}. That is, adjusting the effective sample size would not be an appropriate solution to the problem and  a more flexible non-parametric approach is required to address the various violated assumptions.

\begin{figure}[H]
    \centering
    \begin{subfigure}{\linewidth}
    \centering
        \begin{tabular}{cc}
        \includegraphics[width = 0.3\linewidth, trim=0 0.85in 0in 0, clip]{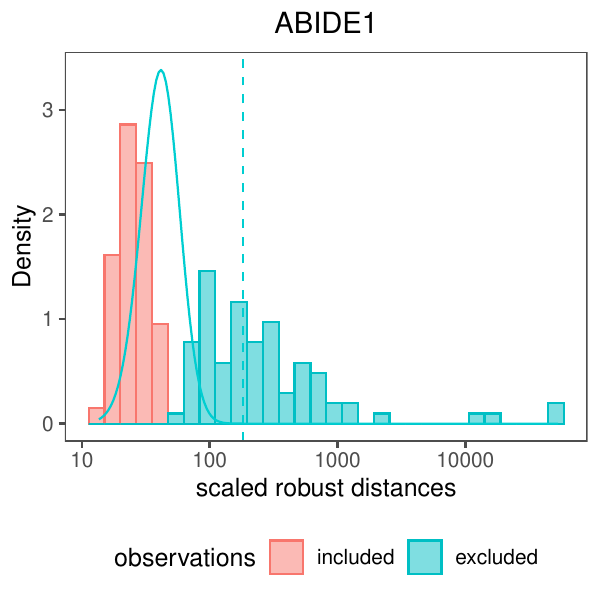} &
        \includegraphics[width = 0.3\linewidth, trim=0 0.85in 0in 0, clip]{image/Figure3_Distributions_of_RD_from_ABIDE2.pdf} \\
        \end{tabular}
        \begin{tabular}{c}
        \includegraphics[width = 0.4\linewidth, trim=0 0in 0in 3.6in, clip]{image/Figure3_Distributions_of_RD_from_ABIDE2.pdf} \\
        \end{tabular}
        \caption{Distribution of RDs of toy fMRI datasets, theoretical distribution and threshold value (dashed) based on the theoretical F distribution (solid line) }
        \hspace{10em}
    \end{subfigure}%
    \hfill
    \hspace{2em}
    \begin{subfigure}{\linewidth}
    \begin{tabular}{c|c}
         \begin{picture}(10,90)\put(0,45){\rotatebox[origin=c]{90}{ABIDE1}}\end{picture} & \includegraphics[width = 0.95\linewidth,trim=0in 0in 0in 0.53in, clip]{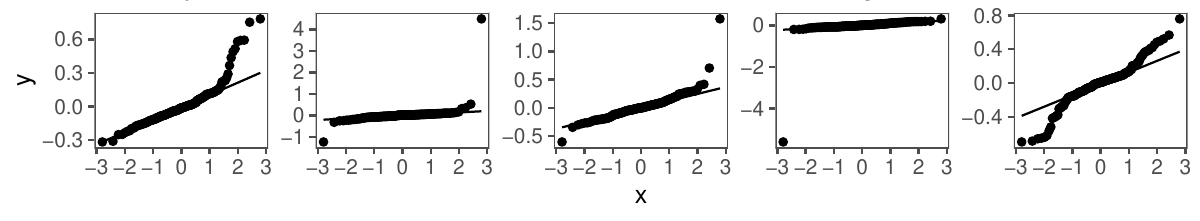} \\
         \hline
         \begin{picture}(10,90)\put(0,45){\rotatebox[origin=c]{90}{ ABIDE2}}\end{picture} &  \includegraphics[width = 0.95\linewidth,trim=0in 0in 0in 0.53in, clip]{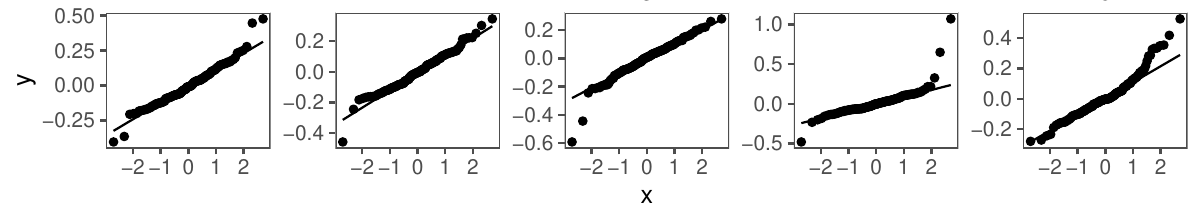}
    \end{tabular}
        \caption{Normal QQ-plots of several components for ABIDE1 (out of 26) and ABIDE2 (out of 5)}
    \end{subfigure}
\caption{\small \textbf{fMRI data violate the Gaussianity assumption, leading to an RD distribution that does not follow the theoretical F distribution.} (A) The dashed-vertical lines represent the $99^{th}$ quantile of the F distribution. RDs of ABIDE1 range from $10$ to over $10000$. Volumes with RD greater than $150$ are labeled as outlier based on the threshold value. Considering the range of the RDs, the threshold value does not indicate a resonable value. Unlike ABIDE1, the range of RDs of ABIDE2 is very short. Given that ABIDE2 is known to be cleaner, the RDs tend to be smaller overall, but the distribution still deviates from the theoretical F distribution, and the $99^{th}$ quantile of the F distribution may be overly aggressive. (B) QQ-plots of several components after applying dimension reduction, showing significant and heterogeneous deviations from Gaussian distribution for both datasets.} 
    \label{fig:hist-ABIDE}
\end{figure}

In addition to violating the independence assumption, fMRI data also violate the Gaussianity assumption. To demonstrate this and the subsequent failure of the theoretical distribution, we employ two ``toy" fMRI sessions based on a single brain slice from two fMRI sessions from the publicly available ABIDE database \citep{di2014ABIDE}. Based on visual inspection of the original fMRI data, the first fMRI session, \textit{ABIDE 1}, is known to be highly contaminated with artifacts, while the second fMRI session, \textit{ABIDE 2}, is relatively free of artifacts. Figure \ref{fig:hist-ABIDE} displays the empirical distribution of RDs for each dataset after dimension reduction, as described in Section \ref{sec:dimred}. Q-Q plots are shown to check the Normality assumption for $5$ of the components. Both datasets show violations of Gaussianity. The Q-Q plots of the selected components highlight that they deviate from Gaussianity, which cannot be explained solely by the presence of outliers. Their distributions are non-Gaussian in different ways, which prevents the application of a common transformation to achieve Normality. As a result, the scaled F distributions clearly fail to fit the distribution of the RDs' excluded observations from both datasets. The empirical distribution of RDs is greater than what would be expected based on the theoretical distribution. The $99^{th}$ quantile of the theoretical F distribution falls within the middle of the distribution of excluded observations for both data. This would lead to many likely non-outlying observations being classified as outliers. The conclusion is that, while Hardin \& Rocke's theoretical result holds for i.i.d. Gaussian data and even moderately correlated Gaussian data, their approach fails when the assumption of Gaussianity is violated, making it inappropriate for fMRI data.

\subsection{Proposed method to determine the distribution of robust distance non-parametically}
\label{sec:bootRD}

Since fMRI data are autocorrelated and non-Gaussian, two key assumptions of the theoretical approach to determining a distribution for RDs of excluded observations are violated. Although typical fMRI data could also violate their third assumption, which is to be from identically distributed data, we seek to satisfy it by applying mean and variance detrending to each independent component. Therefore, a method of determining the distribution of RDs and identifying outliers is required to overcome both independence and Gaussianity violation. Here, we introduce a novel technique for estimating the $(1-\alpha)$ quantile of the distribution of RDs based on univariate outlier imputation and (optionally) a non-parametric bootstrap procedure. The following subsections describe the procedure. Section \ref{sec:uoi} describes a robust univariate outlier identification and imputation procedure used to prevent outliers being masked, where an \textit{empirical} quantile will be calculated after this step. Section \ref{sec:estQ} describes the details of estimating the $(1-\alpha)$ quantile of the distribution of RDs by using a non-parametric bootstrap and the choice of a threshold measure summarizing over bootstrap samples to identify outliers.

\subsubsection{Robust univariate outlier imputation}
\label{sec:uoi}

The existence of outliers in the data alters the distribution of RDs, particularly the upper quantiles. For instance, for a dataset of size $n$, assuming there is only one outlier, bootstrap samples from that dataset would contain that outlier with probability $\left(1- \frac{1}{n} \right)^{n} \to \frac{1}{e} \approx 0.37$ as $n \to \infty$. In fMRI data, there are typically numerous outlying volumes. Prior to estimation of a quantile, we therefore apply univariate outlier imputation to each component of the dimension-reduced dataset. Algorithm \ref{app:ruoi}  describes the steps to obtain imputed data. First, we robustly transform each component to achieve central Normality, as proposed by \cite{raymaekers2021transforming}. The goal of this transformation is to prevent outliers from being masked after power transformations (e.g. Box-Cox or Yeo-Johnson). This transformation aims to achieve Normality for the center of the data distribution while retaining the outlying observations in the tail of the distribution. 

Following the transformation, we identify outliers by using the median absolute deviation (MAD), which is a robust measure of variability for univariate data. For a given observation ${\bf{x}} = \{ {x}_1,{x}_2, ..., {x}_n \}$, the MAD is defined as the median of the absolute deviations from the median of the data $MAD = med(|{\bf{x}} - M|)$ where $M = med({\bf{x}})$. We use a scaling factor of $1.4826$ to make the MAD a consistent estimator of the standard deviation for Gaussian data \citep{rousseeuw1993alternatives}. Observations lying beyond $4$ MADs of the median are considered outliers. To obtain an imputed dataset for each outlying observation, we use the mean of the nearest preceding and following non-outlier observations. We will use this imputed dataset, ${\bf{X}}^{0}_{T \times K}$, to determine the distribution of RDs in the absence of outliers, as described next. 

\subsubsection{Estimation of $(1-\alpha)$ quantile of robust distance}
\label{sec:estQ}

Our goal is to estimate the $(1-\alpha)$ quantile of the distribution of RDs, which will be used to threshold the observed RDs to identify outliers. Here, $\alpha$ represents the proportion of non-outliers that are expected to be labeled as outliers, i.e. false positives. To estimate the $(1-\alpha)$ quantile while accounting for uncertainty, we propose a bootstrap procedure described in Algorithm \ref{alg:boot}. First, we divide observations into included and excluded MCD subsets as described in Section \ref{sec:RD}. We then bootstrap included and excluded observations separately to preserve the MCD structure. For each bootstrap sample, we compute the RD of each observation in the bootstrap sample using the bootstrap MCD mean and the main MCD covariance. We record the $(1-\alpha)$ quantile of the RD in each bootstrap sample. This process is repeated across $B$ bootstrap samples to obtain a bootstrap distribution of the $(1-\alpha)$ quantile of the distribution of RDs. Finally, we can choose a threshold measure among several possible summary statistics from this bootstrap distribution. 

\begin{algorithm}[H]
\DontPrintSemicolon
\SetAlgoLined
\textbf{Input}: univariate outlier imputed data ${\bf{X}}^{0}_{T \times K}$ and $\alpha \in (0,1)$ \;
Define the index sets $S_1$ and $S_2$ for $\textit{included}$ and $\textit{excluded}$ observations of ${\bf{X}}^0$, respectively \;
Compute main MCD covariance ${\hat{\boldsymbol{\Sigma}}}_{MCD}$ from $\{{\bf{x}}^0_{t} : t \in S_1 \}$ where ${\bf{X}}^0_{t}$ is the t-th row of ${\bf{X}}^0$ \;
\For{b = 1,2,.., B}{
  use $F_1 = \frac{1}{n}$ probability mass to sample each of $n$ included observations to create $S_{1}^{(b)}$ \;
  use $F_2 = \frac{1}{n-h}$ probability mass to sample each of $n-h$ excluded observations to create $S_{2}^{(b)}$ \;
  compute MCD estimate of sample mean ${\hat{\boldsymbol{\mu}}}_{MCD}^{(b)}$ from $\{ {\bf{x}}^0_{t} : t \in S^{(b)}_{1} \}$ \;
  compute bootstrap-based RDs of $\{ {\bf{x}}^{0}_t : t \in S^{(b)}_1 \bigcup S^{(b)}_2 \}$ by using  ${\hat{\boldsymbol{\Sigma}}}_{MCD}$ and ${\hat{\boldsymbol{\mu}}}_{MCD}^{(b)}$ \;
  calculate $(1-\alpha)$ quantile estimates, ${\hat{Q}}^{(b)}_{(1-\alpha)}$, from bootstrap-based RDs 
}
\Return the estimated {${\hat{Q}}^{(b)}_{(1-\alpha)}$} for $b=1,2,...,B$
  \caption{Bootstrap-based Estimation of the $(1-\alpha)$ Quantile of the distribution of RD}
  \label{alg:boot}
\end{algorithm}


Depending on the aim of a study, there may be different objectives for outlier detection, including nominal specificity, higher sensitivity, or higher specificity. If the objective is nominal specificity, equivalent to nominal false positive rate of $\alpha$, then the mean or median of the bootstrap distribution could be used to estimate the ($1-\alpha$) quantile of the RD distribution. However, if the objective is high sensitivity to outliers (as in the fMRI scrubbing context), then one could use either a lower bound of a bootstrap confidence interval or the lower bound of a bootstrap CI. On the other hand, if the objective is high specificity to avoid discarding observations that are not true outliers, one could use the lower bound of a bootstrap CI. Hence, we consider two approaches: the empirical quantile and the lower bound (LB) of a bootstrap CI. To compare these two and assess their effects on FPR, we generate $1000$ outlier-free replicates of i.i.d and Gaussian data of size $n = 1000$ and set $\alpha$ to $0.01$. We start by applying Algorithm \ref{alg:boot}, which results in a distribution of the bootstrap-based $99^{th}$ quantile.  

\begin{figure}[H]
    \centering
    \includegraphics[width=0.4\textwidth]{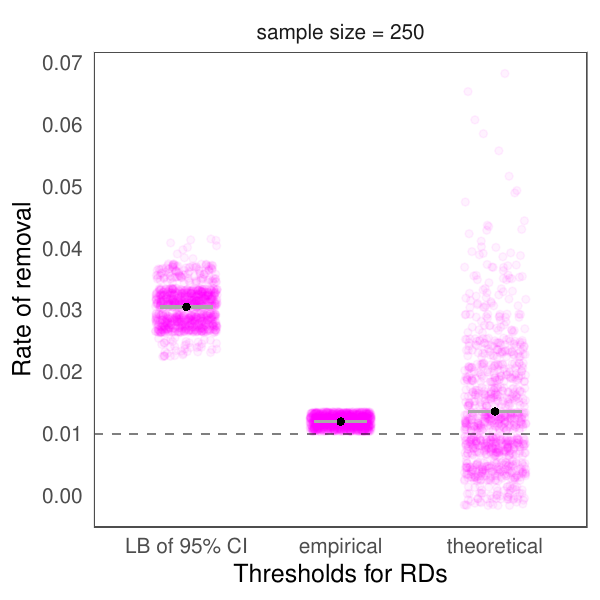}
    \caption{\small \textbf{Rate of removed outliers in simulated standard Normal data, mean across replicates (block dot), and standard error bars of the mean across the replicates (gray).} For fMRI data, it is crucial to identify all outliers, since any artifactual volumes retained can negatively affect the downstream analysis. The empirical threshold generally achieves at least a $1\%$ removal rate while maintaining a removal rate consistently below $2\%$. On the other hand, the LB of the $95\%$ bootstrap CI consistently achieves at least a $1\%$ removal rate, but has an average removal rate of $3\%$. We also show the removal rate of the theoretical threshold, which is around the nominal level of $1\%$ on average (note that the data here are Gaussian, matching the assumptions of this method unlike real fMRI data), but exhibits higher level of variability than the other two threshold approaches. In addition, the empirical threshold is less variable than the LB of the $95\%$ bootstrap CI and theoretical threshold.}
    \label{fig:FPR}
\end{figure}

For the objective of high sensitivity, we consider using the lower bound of a one-sided bootstrap confidence interval (CIs) of the $(1-\alpha)$ quantile as a potential cutoff. Specifically, we consider $97.5\%$ lower one-sided CI ($95\%$ CI LB). The FPR for each replicate is illustrated in Figure \ref{fig:FPR} along with the means across replicates. As seen in Figure \ref{fig:FPR}, this threshold achieves FPR above $0.01$ in all replicates, indicating that it avoids below-nominal FPR. We also consider the $0.99$ empirical quantile as a possibility to reduce the time complexity. The FPR across replicates achieve the strictest type-1 error control, maintaing FPR below $0.02$ in all of the replicates in the simulated example. For comparison, we also consider using the $0.99$ quantile of the theoretical F distribution. The theoretical F distribution is more variable than other cutoff types and often above the nominal rate of $0.01$. The empirical quantile FPR is closer to the nominal rate and are less variable than the theoretical distribution. Considering that this is Gaussian data, the theoretical threshold achieves close to nominal FPR would be worse in fMRI data.

For a given choice of threshold approach, there is variability in the sensitivity to outliers using that approach. That can be due to sampling variability in the estimate and/or variability in the distribution of the data. As a result, a particular choice of threshold may yield outlier sensitivity rates that are higher or lower than desired. It is important to take this into account in order to avoid being overly conservative or overly liberal when identifying outliers, depending on the data context. In the context of fMRI, it is important to identify all outliers, since any artifactual volumes retained can have negative effects on downstream analysis. This is consistent with the approach often adopted in head motion-based scrubbing, where low (stringent) thresholds for head motion are often used in order to remove as much noise as possible, even though some signal is also discarded.

As demonstrated in \cite{pham2023less}, data-driven scrubbing techniques, such as the one proposed here, can effectively reduce noise and retain much less data by better distinguishing signal and noise. However, it remains crucial to perform stringent noise removal is in fRMI analysis. Therefore, we consider a conservative threshold, the empirical threshold. Figure \ref{fig:FPR} shows that the empirical threshold generally achieves removal rates of at least $1\%$, while consistently maintaining a removal rate below $2\%$. On the other hand, the LB of the $95\%$ bootstrap CI consistently achieves at least $1\%$, but has an average removal rate of $3\%$. We also show the removal rate of the theoretical threshold, which is around the nominal level of $1\%$ on average (note that the data used here are Gaussian and match the assumptions of this method, unlike real fMRI data), but exhibits a higher level of variability than the other two threshold approaches. Additionally, the empirical threshold is less variable than the  $95\%$ bootstrap CI and theoretical threshold. Figure \ref{fig:FPR} illustrates that the empirical quantile provides a more sensitive and stable threshold for outliers than the LB of the $95\%$ bootstrap CI. 

\begin{figure}[H]
    \centering
    \includegraphics[width=0.53\textwidth]{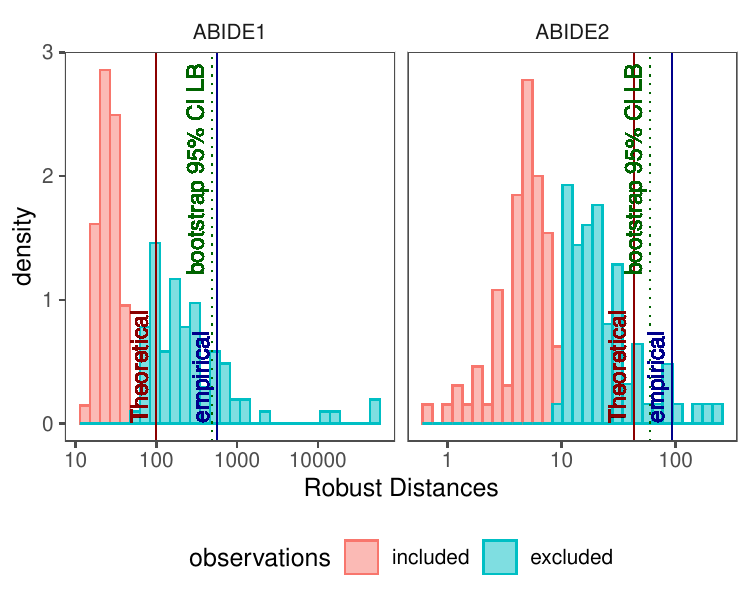}
    \caption{\small \textbf{Distribution of RDs of ABIDE datasets, theoretical threshold and proposed thresholds.} The theoretical quantile performs poorly with more than half of the excluded observations labeled as outliers, which contradicts with the higher sensitivity goal in fMRI scrubbing analysis. By contrast, the empirical and LB of $95\%$ bootstrap CI cutoffs perform well, successfully identifying the long upper tail likely to consists of outliers in both datasets. The bootstrap CI LB method is slightly more stringent than the empirical threshold. \\}
    \label{fig:ABIDE-cutoffs}
\end{figure}

We apply the same approach to both ABIDE toy datasets. As illustrated in Figure \ref{fig:ABIDE-cutoffs}, it is clear that the theoretical quantile performs poorly, with more than half of the excluded observations labeled as outliers. This is particularly bad for ABIDE1, the dataset with more artifacts. By contrast, the empirical and LB of $95\%$ bootstrap CI cutoffs perform well, successfully identifying the long upper tail likely to consist of outliers in both datasets.

\section{Experimental data results}
\label{sec:EDA}

We apply both the empirical and LB of bootstrap CI outlier identification methods to fMRI data and compare their performance with existing methods. This section is organized as follows. In Section \ref{sec:fMRI}, we introduce the fMRI data we employ. We apply our proposed outlier identification procedure to some selected data and present the results on the RDs distribution. In Section \ref{sec:eval}, we compare our proposed approaches with existing scrubbing methods in terms of ther effectiveness on functional connectivity. In Section \ref{sec:spatial}, illustrate the artifacts associated with spatial patterns of our proposed methods. 

\subsection{fMRI datasets}
\label{sec:fMRI}

We employed publicly available data from the Human Connectome Project (HCP) \citep{van2013wu}. This data resource includes resting-state fMRI (rs-fMRI) from $1200$ healthy young adults, and details of the acquisition and processing can be found in \cite{van2013wu} and \cite{glasser2013minimal}. Each subject was scanned over two sessions (REST1 and REST2), and at each session, there were two runs acquired with different phase encoding (LR and RL). $45$ participants were retested (RETEST) with the two sessions and two acquisition directions. Since we compared our approach with the recently developed scrubbing approach \citep{pham2023less}, we employed the same data from $42$ of these subjects. That is, for each subject, there were $8$ sessions. Each of these data included $1200$ volumes acquired every $0.72$ seconds over approximately $15$ minutes. However, the first $15$ volumes were excluded from the analysis to eliminate magnetic field instabilities, resulting in $1185$ volumes per subject. The total number of voxels inside the brain was approximately $200,000$ but varied across subjects. After masking out non-brain areas, a matrix ${\bf{Y}}_{T \times V}$ was created, where $T$ is the acquired volumes for an experiment, and $V$ is the number of voxels in the brain. We employed the \texttt{ciftiTools R} package \citep{pham2022ciftitools} to read, process, and analyze the data.

\begin{figure}[H]
    \centering
    \includegraphics[width=1\textwidth]{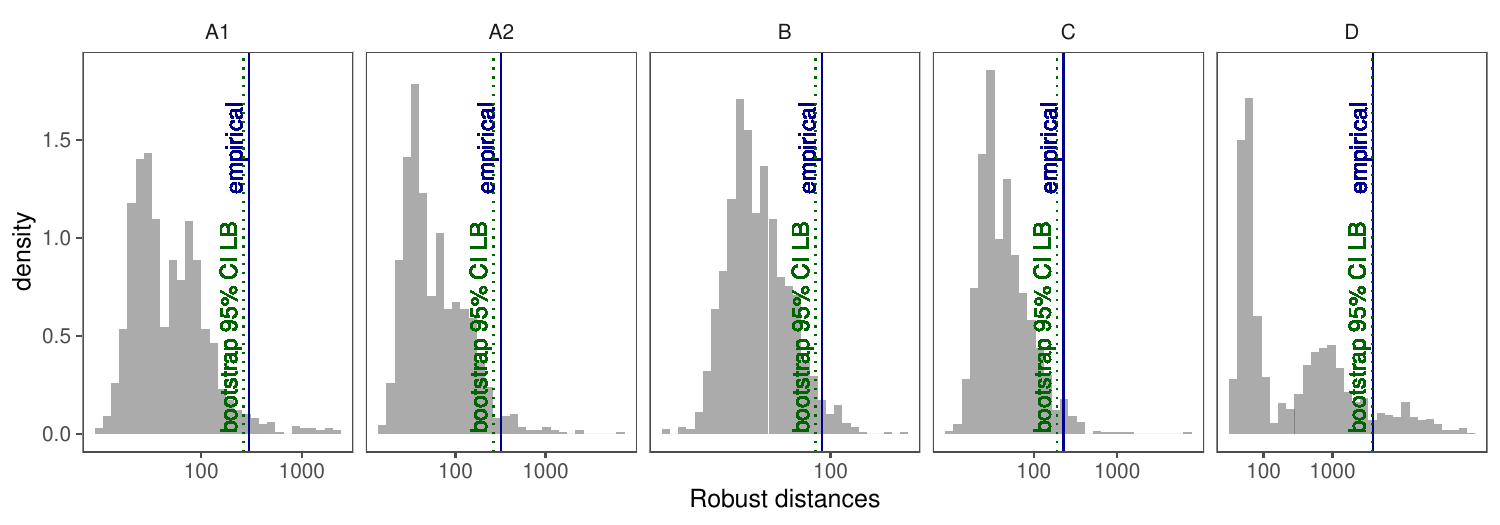}
    \caption{\small \textbf{Distributions of the robust distances for $5$ noisy HCP sessions}. The vertical lines show two estimates of the $0.99$ quantile of RDs: empirical (solid) and bootstrap $95^{th}$ CI LB (dotted). Both threshold values identify quite similar artifactual volumes, with the bootstrap $95^{th}$ CI LB cutoff being slightly stringent. Both approaches seem to threshold the RDs of the data at a reasonable level.\\}
    \label{fig:fMRIRD}
\end{figure}

First, we visualize our two threshold candidates (empirical vs bootstrap $95\%$ CI LB) with RDs of fMRI data. To illustrate our method, we selected $5$ highly noisy sessions shown in Figure \ref{fig:fMRIRD}. The same subject, "A", and three rs-fMRI data from different subjects ("B", "C", and "D") exhibit high levels of noise and artifacts. We refer to these five sessions as \textit{A1}, \textit{A2}, \textit{B}, \textit{C}, and \textit{D}. For each session, we first apply dimension reduction and selection method as described in Section \ref{sec:estQ}. Second, we compute the RDs of each volume, and we obtain thereshold values based on the empirical vs bootstrap $95\%$ CI LB\ref{sec:estQ}. Figure \ref{fig:fMRIRD} shows the empirical distribution of RDs of each fMRI session. Since both cutoff values look very close to each other, we adopt the empirical as a threshold for further analysis to avoid the time complexity that occurs during the bootstrap procedure. 

\subsection{Effect of scrubbing on validity of functional connectivity}
\label{sec:eval}

Here, we compare our proposed outlier detection approach with existing scrubbing techniques for fMRI data, including data-driven scrubbing (e.g., projection scrubbing, DVARS) and motion scrubbing. ``Projection scrubbing'' is based on the same dimension reduction and selection procedure but employs a non-robust distance measure, leverage, and uses an ad-hoc threshold for outliers \citep{pham2023less}. Specifically, artifactual volumes are identified based on having a leverage value greater than $3$ times of the median across all volumes. While this approach has been shown to outperform existing data-driven and hardware-based approaches \citep{pham2023less}, it may suffer from masking and/or swamping effect due to the relationship of leverage with Mahalanobis Distance. 

Another common practice to detect artifactual volumes in fMRI data is motion scrubbing using frame-wise displacement (FD), a measure of subject head motion. This approach aims to detect artifacts caused by head movement \citep{power2012, power2014}. FD is a summary measure based on six rigid body realignment parameters used to align a subject's brain across volumes within an fMRI session. We adopted a lagged and filtered version of FD designed for HCP-style multi-band data (``modified FD"), as described by \cite{pham2023less}. A threshold is applied to these time-specific FD measures to flag high motion volumes. Although this approach is commonly used in fMRI analysis, there is no principled or universally accepted threshold value. By following the most common practice across studies, we use $0.2$ as a FD threshold.

\cite{pham2023less} recently conducted a comprehensive comparison of data-driven scrubbing techniques (projection scrubbing, DVARS) and motion scrubbing. They found that data-driven scrubbing was at least as effective as motion scrubbing for downstream analyses based on functional connectivity (FC), while removing much less data ($\sim3\%$ versus $\sim18\%$ of volumes).  One of the metrics considered by \cite{pham2023less} was the mean absolute change (MAC), proposed as a measure of the impact of scrubbing on validity by \citep{power2014,power2014methods, williams2022advancing}. Broadly, MAC compares two scenarios. In the first scenario, volumes are removed from the data based on a particular scrubbing method. In the second scenario, the same number of volumes are removed randomly. This process is repeated across many random selections. Then, MAC is based on the difference in changes in FC with scrubbing versus random removal. If $\Delta Z_{rsp}$ is the change in FC values for a given scan between scrubbing and random scrubbing across $Q$ permutations, then MAC is calculated as follows:

\begin{align}
    \text{MAC} = \frac{1}{SP}  \mathlarger{\sum}_{s=1}^{S} \mathlarger{\sum}_{p=1}^{P} \Bigg| \frac{1}{R} \mathlarger{\sum}_{r=1}^{R} \Delta z_{rsp} \Bigg| , \label{eqn:MAC}
\end{align}

where $S$ is the number of subjects, $P$ is the unique FC pairs, and $R$ is the random scrubbing permutation. At a fixed censoring rate, higher MAC values are an indication of improved validity of FC. Because MAC is expected to increase with censoring rate, to compare these values successfully the censoring rates must be fixed.

\begin{figure}[H]
    \centering
    \includegraphics[width=0.9\textwidth]{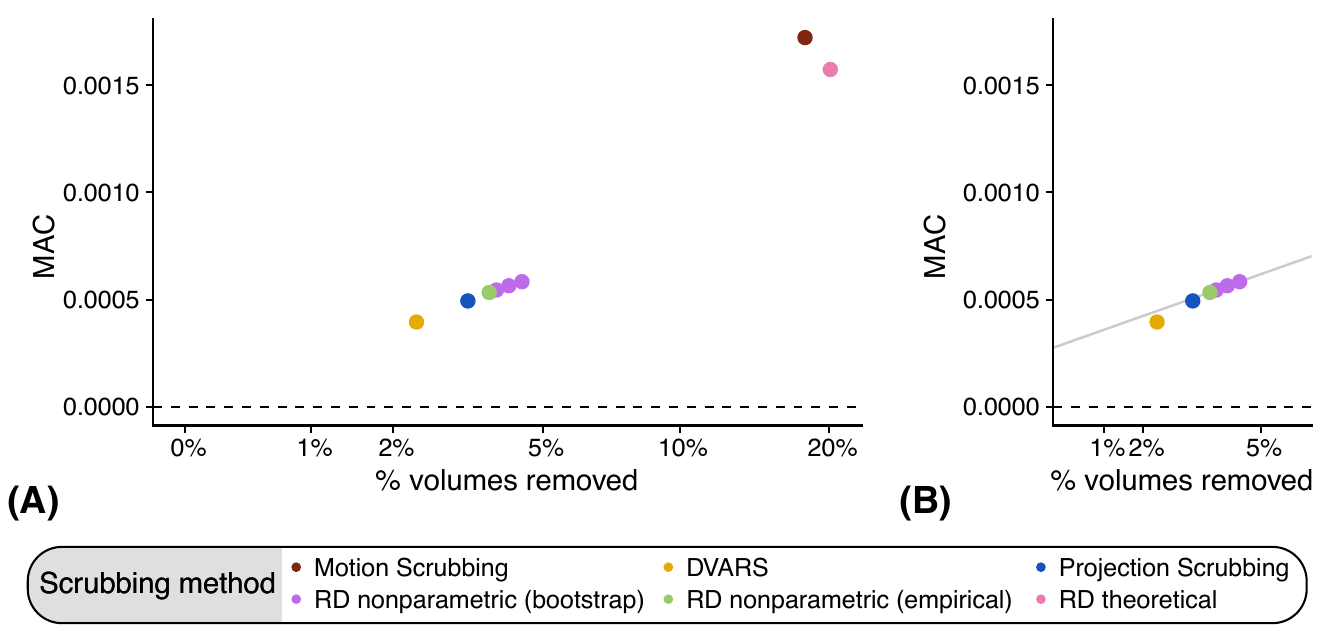}
    \caption{\textbf{Effect of different scrubbing methods on MAC.}  The x-axis shows censoring rate ($\%$ frames removed) and the y\-axis shows MAC. Since MAC is expected to increase with higher censoring rate, it is compared across a fixed censoring rate. (A) Comparisons of aforementioned scrubbing methods. This shows that scrubbing (modified FD, dark red) and RD with the theoretical cutoff (RD theoretical, pink) result in very high censoring rates near $20\%$. Comparing these two, modified FD seems to be better than RD theoretical based on MAC. (B) Comparisons of scrubbing methods producing closer MAC measures. Our proposed method is displayed in purple for three different lower bound of bootstrap CI ($50\%$, $80\%$, $95\%$) and in green with the empirical cutoff. The reference line is displayed to compare the methods. Our proposed method flags slightly more volumes than projection scrubbing, suggesting it may help to avoid masking of smaller outliers that may occur with projection scrubbing due to the use of a non-robust distance measure.  Our proposed method may also be slightly better in terms of MAC.  Both projection scrubbing and our proposed robust method perform better than DVARS in terms of MAC.}
    \label{fig:MAC}
\end{figure}

In Figure \ref{fig:MAC}, we illustrate the effects of various scrubbing approaches on the effectiveness of FC by using the MAC metric. Note that since MAC is expected to increase with a higher censoring rate, it should only be compared across a fixed censoring rate. The left-hand plot displays all methods and shows that motion scrubbing (dark red) and RD with the theoretical cutoff (RD theoretical, pink) result in very high censoring rates near $20\%$.  Comparing these two, modified FD seems to be better than the theoretical based on MAC. The right-hand plot shows the same results on a narrower x-axis scale to facilitate comparison across the remaining methods, including our proposed approach. Our proposed method is displayed in purple for three different bootstrap-based cutoffs ($50\%$, $80\%$, $95\%$) and in green with the empirical cutoff. A reference line is plotted to facilitate comparison with ICA projection scrubbing and DVARS. This shows that our proposed method flags slightly more volumes than projection scrubbing, suggesting it may help to avoid masking of smaller outliers that may occur with projection scrubbing due to the use of a non-robust distance measure. Our proposed method may also be slightly better in terms of MAC. Both projection scrubbing and our proposed robust method perform better than DVARS in terms of MAC.

\subsection{Spatial patterns of the artifactual volumes}
\label{sec:spatial}

Finally, we visualize the artifacts associated with spatial patterns of RD-based artifactual volumes for each high RD volume. Recall that after performing ICA to obtain ${\bf{Y}} = {\bf{AS}} + {\bf{E}}$, $\bf{X}$ is obtained by selecting a subset of high-kurtosis columns of $\bf{A}$. Let ${\bf{A}}^*$ and ${\bf{S}}^*$ contain only the selected rows and columns of ${\bf{A}}$ and ${\bf{S}}$. For each outlying volume $t$, an image of artifact intensity can be obtained by multiplying the $t$-th row of ${\bf{A}}^*$ by ${\bf{S}}^*$. SSince it is not possible to display each of these volumes due to space constraints, we visualize the average across all outlying volumes, which gives an overall measure of artifact intensity at every voxel of the brain.

\begin{figure}[H]
    \centering
    \includegraphics[width=0.84\textwidth]{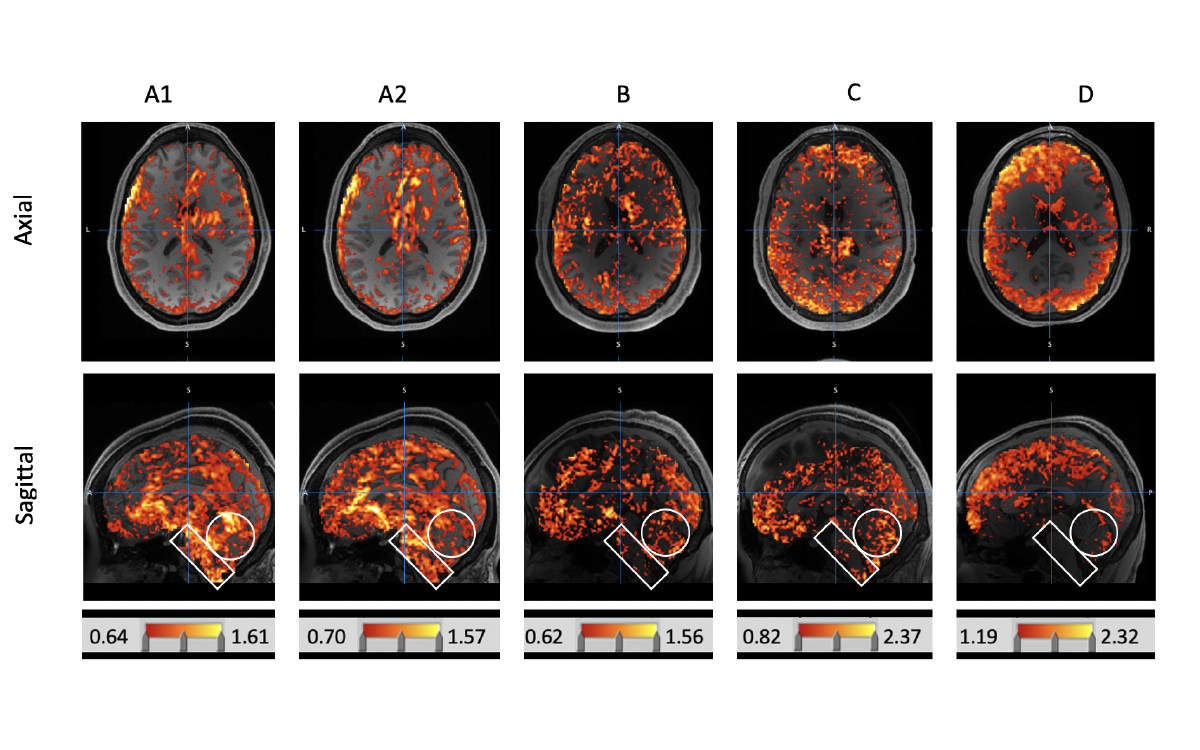}
    \caption{\small \textbf{Average spatial patterns of artifact intensity for five high-noise sessions.} The bottom color scales indicate the subject-specific signal magnitudes. Note that the images are thresholded at the mean to aid in visualization. The white rectangular on each sagittal view shows the spinal cord of the brain. The white circle on each sagittal view highlights the cerebellum of the brain and identified artifacts that area are indication of head motion sourced artifact.  \\}
    \label{fig:brain}
\end{figure}

Using the same five high-noise sessions shown in Figure \ref{fig:fMRIRD}, we display two orthogonal views of brain images (slices) acquired from each subject, containing artifact intensities, in Figure \ref{fig:brain}. The intensities are lower when the color is more red and higher when the color is more yellow. The axial views of each subject show a ring of intensity on the outer edge of the brain, which is often caused by head movements during data acquisition, leading to mislocalization of signals. The sagittal view of each image demonstrates high artifact intensity around the spinal cord (white rectangle), which is another indication of head movement-related artifacts. It is interesting to see similar artifact patterns across different sessions of the same subject (A1 and A2). In contrast to the patterns of artifacts seen in sessions \textit{A1}, \textit{A2}, \textit{B}, and \textit{C}, the sagittal view of session \textit{D} illustrates tissue activation around the edge of the cerebellum (white circle on the sagittal views), which could be an indication of pulsatile artifact from blood being pumped in brain blood vessels with each heart beat. In addition to these spatial differences, the subject-based color scales highlight the signal changes across subjects. For instance, session \textit{D} exhibits more intense artifacts compared to subject \textit{A}. This agrees with the higher magnitude of RDs of session D compared to the other sessions, as seen in Figure \ref{fig:fMRIRD}.

\section{Discussion and future work}
\label{sec:discussion}
We have proposed a robust non-parametric multivariate outlier detection method that is applicable to fMRI data, which exhibits markedly non-Gaussian and non-independent features. We have compared our proposed approach with an established robust but parametric outlier detection method based on the theoretical F distribution of RD established by \cite{hardin2005}. While this method is suitable for identical, independent, and normally distributed data, it cannot be applied to fMRI data \citep{hardin2005,filzmoser2008, cerioli2010multivariate, mejia2017pca}. Violation of the assumptions causes the distribution of RD to deviate from the theoretical distribution, which can lead to an increase in false positive rate or a decrease in sensitivity to outliers, depending on the nature of the violations. Prior work has attempted to reduce the distributional deviation between the theoretical and empirical fit through ad hoc methods. Our proposed non-parametric approach instead uses univariate outlier imputation, combined with a bootstrap procedure to estimate relevant summary statistics of the true distribution of RD.

The proposed method also offers a statistically principled alternative to motion scrubbing, which is known to result in high rates of volume censoring, to identify artifactual volumes in fMRI data. Motion scrubbing often employs an ad-hoc but strict threshold value that could cause more than half of the volumes to be flagged as outliers, excluding potentially valuable data. Moreover, motion scrubbing can only detect artifacts coinciding with head motion and may therefore miss other types of artifacts. Our proposed method also improves upon existing data-driven scrubbing techniques such as projection scrubbing \citep{pham2023less} and DVARS \citep{smyser2010longitudinal, afyouni2018insight}, as a formal robust outlier detection framework that is appropriate for fMRI data.

Although the proposed method outperforms existing methods, it has several limitations. First, we use neighboring observations of each column of data. To resolve this, we could use a more generalized outlier imputation approach, which imputes better by using information from other columns. Incorporating multiple imputation into the bootstrap samples could also provide better results. Second, we employ a robust transformation prior to individual components to reduce skew prior to univariate outlier detection. However, components often exhibit other non-Gaussian features such as heavy tails. In the future, we plan to develop a more flexible robust transformation that is appropriate for a wider range of non-Gaussian distributions.

Additional validation studies are required to better understand the performance of our proposed method. First, while the proposed method's excellent performance is demonstrated in simulated data, some additional tests are still needed. For example, the method could be tested on non-normal and outlier-contaminated simulated data. Second, the method has been applied to resting-state fMRI data, which contains task-free signals. However, the method could be applied and tested on task-related fMRI data by regressing out the task-related signals. Third, we employed data from one subject with two separate imaging sessions that showed very similar patterns for artifacts. This attribute could be studied for more subjects to explore whether there is recognizable inter-individual variability in the spatial patterns of artifacts in fMRI datasets. Analyzing more subjects would also allow us to better quantify the performance of our method compared with existing data-driven and software-based scrubbing techniques. Fourth, identification of eye movement-related artifacts is still an important topic of ongoing research. Future work may apply our proposed approach to data where eye tracking is available to see if it is capable of detecting blinks and eye movements better than existing scrubbing techniques, since eye movements can reduce the quality

\bibliography{mybib.bib}

\begin{thebibliography}{31}
\providecommand{\natexlab}[1]{#1}
\providecommand{\url}[1]{\texttt{#1}}
\expandafter\ifx\csname urlstyle\endcsname\relax
  \providecommand{\doi}[1]{doi: #1}\else
  \providecommand{\doi}{doi: \begingroup \urlstyle{rm}\Url}\fi

\bibitem[Afyouni and Nichols(2018)]{afyouni2018insight}
S.~Afyouni and T.~E. Nichols.
\newblock Insight and inference for dvars.
\newblock \emph{Neuroimage}, 172:\penalty0 291--312, 2018.

\bibitem[Afyouni et~al.(2019)Afyouni, Smith, and Nichols]{afyouni2019effective}
S.~Afyouni, S.~M. Smith, and T.~E. Nichols.
\newblock Effective degrees of freedom of the pearson's correlation coefficient
  under autocorrelation.
\newblock \emph{NeuroImage}, 199:\penalty0 609--625, 2019.

\bibitem[Beauchamp et~al.(2003)Beauchamp, Lee, Haxby, and
  Martin]{beauchamp2003fmri}
M.~S. Beauchamp, K.~E. Lee, J.~V. Haxby, and A.~Martin.
\newblock Fmri responses to video and point-light displays of moving humans and
  manipulable objects.
\newblock \emph{Journal of cognitive neuroscience}, 15\penalty0 (7):\penalty0
  991--1001, 2003.

\bibitem[Cerioli(2010)]{cerioli2010multivariate}
A.~Cerioli.
\newblock Multivariate outlier detection with high-breakdown estimators.
\newblock \emph{Journal of the American Statistical Association}, 105\penalty0
  (489):\penalty0 147--156, 2010.

\bibitem[Di~Martino et~al.(2014)Di~Martino, Yan, Li, Denio, Castellanos,
  Alaerts, Anderson, Assaf, Bookheimer, Dapretto, et~al.]{di2014ABIDE}
A.~Di~Martino, C.-G. Yan, Q.~Li, E.~Denio, F.~X. Castellanos, K.~Alaerts, J.~S.
  Anderson, M.~Assaf, S.~Y. Bookheimer, M.~Dapretto, et~al.
\newblock The autism brain imaging data exchange: towards a large-scale
  evaluation of the intrinsic brain architecture in autism.
\newblock \emph{Molecular psychiatry}, 19\penalty0 (6):\penalty0 659--667,
  2014.

\bibitem[Filzmoser et~al.(2008)Filzmoser, Maronna, and Werner]{filzmoser2008}
P.~Filzmoser, R.~Maronna, and M.~Werner.
\newblock Outlier identification in high dimensions.
\newblock \emph{Computational statistics \& data analysis}, 52\penalty0
  (3):\penalty0 1694--1711, 2008.

\bibitem[Frank et~al.(2001)Frank, Buxton, and Wong]{frank2001estimation}
L.~R. Frank, R.~B. Buxton, and E.~C. Wong.
\newblock Estimation of respiration-induced noise fluctuations from
  undersampled multislice fmri data.
\newblock \emph{Magnetic Resonance in Medicine: An Official Journal of the
  International Society for Magnetic Resonance in Medicine}, 45\penalty0
  (4):\penalty0 635--644, 2001.

\bibitem[Friston et~al.(1996)Friston, Williams, Howard, Frackowiak, and
  Turner]{friston1996movement}
K.~J. Friston, S.~Williams, R.~Howard, R.~S. Frackowiak, and R.~Turner.
\newblock Movement-related effects in f{MRI} time-series.
\newblock \emph{Magnetic resonance in medicine}, 35\penalty0 (3):\penalty0
  346--355, 1996.

\bibitem[Glasser et~al.(2013)Glasser, Sotiropoulos, Wilson, Coalson, Fischl,
  Andersson, Xu, Jbabdi, Webster, Polimeni, et~al.]{glasser2013minimal}
M.~F. Glasser, S.~N. Sotiropoulos, J.~A. Wilson, T.~S. Coalson, B.~Fischl,
  J.~L. Andersson, J.~Xu, S.~Jbabdi, M.~Webster, J.~R. Polimeni, et~al.
\newblock The minimal preprocessing pipelines for the human connectome project.
\newblock \emph{Neuroimage}, 80:\penalty0 105--124, 2013.

\bibitem[Griffanti et~al.(2014)Griffanti, Salimi-Khorshidi, Beckmann, Auerbach,
  Douaud, Sexton, Zsoldos, Ebmeier, Filippini, Mackay, et~al.]{griffanti2014}
L.~Griffanti, G.~Salimi-Khorshidi, C.~F. Beckmann, E.~J. Auerbach, G.~Douaud,
  C.~E. Sexton, E.~Zsoldos, K.~P. Ebmeier, N.~Filippini, C.~E. Mackay, et~al.
\newblock Ica-based artefact removal and accelerated fmri acquisition for
  improved resting state network imaging.
\newblock \emph{Neuroimage}, 95:\penalty0 232--247, 2014.

\bibitem[Hardin and Rocke(2005)]{hardin2005}
J.~Hardin and D.~M. Rocke.
\newblock The distribution of robust distances.
\newblock \emph{Journal of Computational and Graphical Statistics}, 14\penalty0
  (4):\penalty0 928--946, 2005.

\bibitem[Kr{\"u}ger and Glover(2001)]{kruger2001physiological}
G.~Kr{\"u}ger and G.~H. Glover.
\newblock Physiological noise in oxygenation-sensitive magnetic resonance
  imaging.
\newblock \emph{Magnetic Resonance in Medicine: An Official Journal of the
  International Society for Magnetic Resonance in Medicine}, 46\penalty0
  (4):\penalty0 631--637, 2001.

\bibitem[Lindquist(2008)]{lindquist2008statistical}
M.~A. Lindquist.
\newblock The statistical analysis of f{MRI} data.
\newblock 2008.

\bibitem[Lopuhaa and Rousseeuw(1991)]{lopuhaa1991}
H.~P. Lopuhaa and P.~J. Rousseeuw.
\newblock Breakdown points of affine equivariant estimators of multivariate
  location and covariance matrices.
\newblock \emph{The Annals of Statistics}, pages 229--248, 1991.

\bibitem[Mahalanobis(1936)]{mahalanobis1936generalised}
P.~C. Mahalanobis.
\newblock On the generalised distance in statistics.
\newblock In \emph{Proceedings of the national Institute of Science of India},
  volume~12, pages 49--55, 1936.

\bibitem[Maronna and Zamar(2002)]{maronna2002}
R.~A. Maronna and R.~H. Zamar.
\newblock Robust estimates of location and dispersion for high-dimensional
  datasets.
\newblock \emph{Technometrics}, 44\penalty0 (4):\penalty0 307--317, 2002.

\bibitem[Mejia et~al.(2017)Mejia, Nebel, Eloyan, Caffo, and
  Lindquist]{mejia2017pca}
A.~F. Mejia, M.~B. Nebel, A.~Eloyan, B.~Caffo, and M.~A. Lindquist.
\newblock Pca leverage: outlier detection for high-dimensional functional
  magnetic resonance imaging data.
\newblock \emph{Biostatistics}, 18\penalty0 (3):\penalty0 521--536, 2017.

\bibitem[Parlak et~al.(2022)Parlak, Pham, Spencer, Welsh, and
  Mejia]{Parlak2022-sz}
F.~Parlak, D.~D. Pham, D.~A. Spencer, R.~C. Welsh, and A.~F. Mejia.
\newblock Sources of residual autocorrelation in multiband task {fMRI} and
  strategies for effective mitigation.
\newblock \emph{Front. Neurosci.}, 16:\penalty0 1051424, 2022.

\bibitem[Pham et~al.(2023)Pham, McDonald, Ding, Nebel, and Mejia]{pham2023less}
D.~Pham, D.~J. McDonald, L.~Ding, M.~B. Nebel, and A.~F. Mejia.
\newblock Less is more: balancing noise reduction and data retention in fmri
  with data-driven scrubbing.
\newblock \emph{NeuroImage}, 270:\penalty0 119972, 2023.

\bibitem[Pham et~al.()Pham, Muschelli, and Mejia]{pham2022ciftitools}
D.~D. Pham, J.~Muschelli, and A.~F. Mejia.
\newblock ciftitools: A package for reading, writing, visualizing, and
  manipulating cifti files in r.
\newblock \emph{NeuroImage}, 250:\penalty0 118877.

\bibitem[Power et~al.(2012)Power, Barnes, Snyder, Schlaggar, and
  Petersen]{power2012}
J.~D. Power, K.~A. Barnes, A.~Z. Snyder, B.~L. Schlaggar, and S.~E. Petersen.
\newblock Spurious but systematic correlations in functional connectivity mri
  networks arise from subject motion.
\newblock \emph{Neuroimage}, 59\penalty0 (3):\penalty0 2142--2154, 2012.

\bibitem[Power et~al.(2014{\natexlab{a}})Power, Mitra, Laumann, Snyder,
  Schlaggar, and Petersen]{power2014}
J.~D. Power, A.~Mitra, T.~O. Laumann, A.~Z. Snyder, B.~L. Schlaggar, and S.~E.
  Petersen.
\newblock Methods to detect, characterize, and remove motion artifact in
  resting state fmri.
\newblock \emph{Neuroimage}, 84:\penalty0 320--341, 2014{\natexlab{a}}.

\bibitem[Power et~al.(2014{\natexlab{b}})Power, Mitra, Laumann, Snyder,
  Schlaggar, and Petersen]{power2014methods}
J.~D. Power, A.~Mitra, T.~O. Laumann, A.~Z. Snyder, B.~L. Schlaggar, and S.~E.
  Petersen.
\newblock Methods to detect, characterize, and remove motion artifact in
  resting state fmri.
\newblock \emph{Neuroimage}, 84:\penalty0 320--341, 2014{\natexlab{b}}.

\bibitem[Pruim et~al.(2015)Pruim, Mennes, van Rooij, Llera, Buitelaar, and
  Beckmann]{pruim2015}
R.~H. Pruim, M.~Mennes, D.~van Rooij, A.~Llera, J.~K. Buitelaar, and C.~F.
  Beckmann.
\newblock Ica-aroma: A robust ica-based strategy for removing motion artifacts
  from fmri data.
\newblock \emph{Neuroimage}, 112:\penalty0 267--277, 2015.

\bibitem[Raymaekers and Rousseeuw(2021)]{raymaekers2021transforming}
J.~Raymaekers and P.~J. Rousseeuw.
\newblock Transforming variables to central normality.
\newblock \emph{Machine Learning}, pages 1--23, 2021.

\bibitem[Rousseeuw and Croux(1993)]{rousseeuw1993alternatives}
P.~J. Rousseeuw and C.~Croux.
\newblock Alternatives to the median absolute deviation.
\newblock \emph{Journal of the American Statistical association}, 88\penalty0
  (424):\penalty0 1273--1283, 1993.

\bibitem[Rousseeuw and Driessen(1999)]{rousseeuw1999}
P.~J. Rousseeuw and K.~V. Driessen.
\newblock A fast algorithm for the minimum covariance determinant estimator.
\newblock \emph{Technometrics}, 41\penalty0 (3):\penalty0 212--223, 1999.

\bibitem[Rousseeuw and Van~Zomeren(1990)]{rousseeuw1990unmasking}
P.~J. Rousseeuw and B.~C. Van~Zomeren.
\newblock Unmasking multivariate outliers and leverage points.
\newblock \emph{Journal of the American Statistical association}, 85\penalty0
  (411):\penalty0 633--639, 1990.

\bibitem[Smyser et~al.(2010)Smyser, Inder, Shimony, Hill, Degnan, Snyder, and
  Neil]{smyser2010longitudinal}
C.~D. Smyser, T.~E. Inder, J.~S. Shimony, J.~E. Hill, A.~J. Degnan, A.~Z.
  Snyder, and J.~J. Neil.
\newblock Longitudinal analysis of neural network development in preterm
  infants.
\newblock \emph{Cerebral cortex}, 20\penalty0 (12):\penalty0 2852--2862, 2010.

\bibitem[Van~Essen et~al.(2013)Van~Essen, Smith, Barch, Behrens, Yacoub,
  Ugurbil, Consortium, et~al.]{van2013wu}
D.~C. Van~Essen, S.~M. Smith, D.~M. Barch, T.~E. Behrens, E.~Yacoub,
  K.~Ugurbil, W.-M.~H. Consortium, et~al.
\newblock The wu-minn human connectome project: an overview.
\newblock \emph{Neuroimage}, 80:\penalty0 62--79, 2013.

\bibitem[Williams et~al.(2022)Williams, Tubiolo, Luceno, and
  Van~Snellenberg]{williams2022advancing}
J.~C. Williams, P.~N. Tubiolo, J.~R. Luceno, and J.~X. Van~Snellenberg.
\newblock Advancing motion denoising of multiband resting-state functional
  connectivity fmri data.
\newblock \emph{NeuroImage}, 249:\penalty0 118907, 2022.

\end{thebibliography}

\appendix

\renewcommand\thefigure{\thesection.\arabic{figure}}
\setcounter{figure}{0}

\newpage
\setcounter{page}{1}

\section{Robust univariate outlier imputation algorithm}\label{app:ruoi}

\begin{algorithm}[H]
\DontPrintSemicolon
\SetAlgoLined
\textbf{Input}: ${\bf{X}}_{T \times K}$ (dimension reduced, detrended $\&$ selected fMRI data) \;
Initialize ${\bf{X}}^{0}_{T \times K}$ $\leftarrow$ ${\bf{X}}_{T \times K}$ \;
\For{k = 1,2,.., K}{
  ${\bf{x}}_k$ k-th column vector \;
  Robustly transform ${\bf{x}}_k$ to be ${\bf{x}}^{'}_k$
  and $M_{k} = med({\bf{x}}^{'}_k)$\;
  Calculate $\text{MAD}_k = med(|{\bf{x}}^{'}_k - M_{k})|)$ \;
  Define $\mathcal{T}_k$ = $\left\{ { t : x^{'}_{t,k}} \notin M_{k} \pm [4 \cdot (1.4826 \cdot \text{MAD}_k)] \right\}$ the set of univariate outlier indices \;
  \For{${t} \in \mathcal{T}_k$}{
    ${x}^{0}_{t,k} \leftarrow \text{mean} \left\{ {x}^{'}_{t-a,k}, {x}^{'}_{t+b,k} \right\}$ where $a = \min \{1,2,..., t-1 : x^{'}_{t-a,k} \notin \mathcal{T}_k \}$ and $b = \max \{1,2,..., T-t : x^{'}_{t+b,k} \notin \mathcal{T}_k \}$. The corresponding ${x}^{'}_{t-a,k}$ and ${x}^{'}_{t+b,k}$ will be dropped if either $a$ or $b$ is null.
  }
}
\Return{${\bf{X}}^{0}_{T \times K}$} the imputed data matrix
\caption{Univariate Outlier Imputation}
\end{algorithm}

\end{document}